\documentclass[9pt,twocolumn,twoside]{osajnl}

\journal{ol} 

\setboolean{shortarticle}{true}

\title{ Force sensing beyond standard quantum limit with optomechanical `soft' mode induced by non-linear interaction }

\author[1]{U. Satya Sainadh}
\author[2,*]{M. Anil Kumar}

\affil[1]{Solid State Institute, Technion, Israel Institute of Technology, Haifa-3200003, Israel.}
\affil[2]{Central University of Tamilnadu, Neelakudy, Tamil Nadu 610005, India}

\affil[*]{Corresponding author:anilksp.m@gmail.com}

\ociscodes{(120.4880) Optomechanics; (120.5050) Phase measurement;(280.4788) Optical sensing and sensors}

\begin{abstract}
We consider an optomechanical (OM) system that consists of a mechanical and an optical mode interacting  through a linear and quadratic  OM dispersive couplings. The system is operated in unresolved side band limit with a high quality factor mechanical resonator. Such a system acts as parametrically driven oscillator, giving access to an intensity assisted tunability of the spring constant. This enables the operation of OM system in its 'soft mode' wherein the mechanical spring softens and responds with a lower resonance frequency. We show that this soft mode can be exploited to non-linearize backaction noise which yields higher  force sensitivity beyond the conventional standard quantum limit.  
\end{abstract}

\setboolean{displaycopyright}{true}
\begin{document}

\maketitle

\par The scientific triumph of gravitational wave detection has pushed the  detection limits of classical force sensing. Classical force sensing with dispersive optomechanical (OM) system has a long history. To realize high sensitive force measurements, many approaches have been studied to reduce quantum noise in the system { \cite{cavesprd,bondurant,tsang,heurs,buchmann,mika}}.
The sensitivity of force measurements are limited by two types of quantum noise: the shot noise of the laser beam at the detection port and the radiation pressure backaction noise introduced by the oscillator \cite{clerkreview}. An optimal trade-off between these noises induces a lower bound for force detection sensitivity, which is the so-called standard quantum limit (SQL) \cite{braginsky,caves}.

However, the SQL is not a fundamental quantum limit. Various schemes involving dispersive interactions were proposed to overcome the SQL in force measurements, such as frequency dependent squeezing of the input beam \cite{jaekel}, variational measurement \cite{vyatchanin}, two-mode OM system \cite{xu} and coherent quantum noise cancellation (CQNC) \cite{tsang,heurs,meystre,vitali} etc.  
On the other hand,  a free particle coupled through a dissipative interaction with an OM system was shown to surpass the SQL \cite{sql}.
 
\par  In the current letter, we propose a new strategy to improve force sensing by surpassing SQL using the soft mode of an OM system in the unresolved side band limit. The inclusion of non-linear dispersive interactions like the quadratic OM coupling (QOC) in addition to the linear OM coupling (LOC), makes the system act like a parametrically driven oscillator. This gives us a handle on the behaviour of the intensity driven mechanical oscillator to be either in a 'stiffer' or in 'softer' mode depending on its new effective spring constant \cite{anilkumar}. The soft mode modifies the quantum backaction noise to a non-linear function of the input power leading to force sensitivities that can surpass SQL. 

\begin{figure}[t]
    \centering
    \includegraphics[scale=0.65]{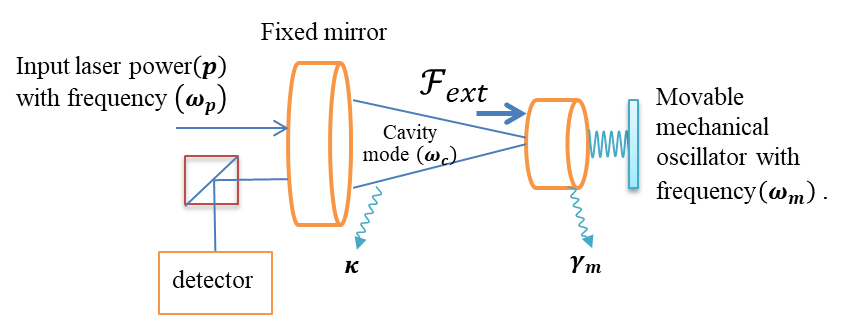}
    \caption{Schematic of an OM system with a cavity frequency $\omega_c$ and line width $\kappa$, interacting with a mechanical oscillator of frequency $\omega_m$ with a decay rate $\gamma_m$ via linear $g_{_{l}}$ and quadratic $g_{_{q}}$ dispersive OM couplings. A classical external force $\mathcal{F}_{_{ext}}$ is acting on the mechanical oscillator. }
    \label{fig:schematic}
\end{figure}
 As shown in the schematic Fig. 1, we consider an OM system and the system's Hamiltonian in the frame of the laser field is,
 \small
 \begin{equation}
 H=\hbar\mathbf{\Delta c^\dagger c}+\frac{\hbar \omega_m }{2}\mathbf{(x^2+p^2)}+\hbar g_{_{l}} \mathbf{c^\dagger c x}+\hbar g_{_{q}} \mathbf{c^\dagger c x^2}+i\hbar\varepsilon\mathbf{(c^\dagger-c)}. \label{a1}
 \end{equation}\normalsize
where $\Delta=\omega_c-\omega_p$ is the detuning of the driving laser frequency with respect to the cavity resonance frequency, $\mathbf{c}$ is the cavity mode's annhilation operator, $\mathbf{x}$ and $\mathbf{p}$ are the normalized position and momentum operators of the mechanical oscillator with $[\mathbf{x,p}]=i$. The first two terms describe the free energy of the cavity and mechanical oscillator and the next two terms describe the OM interaction through LOC ($g_{_{l}}$) and QOC ($g_{_{q}}$). The last term accounts for the pump field amplitude $\epsilon=\sqrt{\frac{2\mathcal{P}\kappa}{\hbar \omega_p}}$, where $\mathcal{P}$ is the input laser power and $\kappa$ as the cavity decay rate.

\par In order to fully describe the dynamics of the system, we write the quantum Langevin equations using { $\frac{d\mathcal{A}}{dt}=\frac{i}{\hbar} [H,\mathcal{A}]$} + dissipation terms.  These  equations are analysed with linear analysis of the fluctuations and dissipation processes affecting the optical and mechanical modes. We expand the time-dependent variables $\mathcal{O}(t)$ around their steady-state values $\mathcal{O}_s$ linearly in fluctuations as $\mathcal{O}(t)=\mathcal{O}_s+\lambda \delta \mathcal{O}(t)$. This gives us 
\small
\begin{eqnarray}
   x_s= \frac{-g_{_{l}}|c_s|^2}{\omega_m+2 g_{_{q}}|c_s|^2}, \qquad 
    c_s  =\frac{\varepsilon}{ \kappa+i\left( \Delta+g_{_{l}}x_s +g_{_{q}}   (x^2)_s \right)}. \label{a3}
 \end{eqnarray}\normalsize 
\par The system is studied under a Markovian evolution with delta correlated noises in both mechanical ($\xi(t)$) and optical modes ($c_{in}(t)$) \cite{aspelmeyer}. The Langevin equations for the fluctuations can be written in a compact form as
$\dot{u}(t)=M u(t)+\nu(t)$ with column vector of fluctuations in the system being $u^T= \left(\begin{matrix} \delta x,&\delta p,&\delta X,&\delta P \end{matrix}\right)$ and column vector of noise being $\nu^T=\left(\begin{matrix}0,&\xi(t),&\sqrt{2\kappa}\delta X_{in},& \sqrt{2\kappa}\delta P_{in}\end{matrix}\right)$ where $\delta X=\frac{\delta c+\delta c^\dagger}{\sqrt{2}}$, $\delta P=\frac{\delta c-\delta c^\dagger}{\sqrt{2}i}$ with  $\delta X_{in}$ and $\delta P_{in}$ are their corresponding noises.  The matrix $M$ is given by
\small
\begin{eqnarray}M=\left(\begin{matrix} 0&\omega_m &0&0\\ -\tilde{\omega}_m& -\gamma_m& 
-\tilde{G} X_s&-\tilde{G} P_s\\ \tilde{G} P_s& 0&-\kappa  & \tilde{\Delta}\\ -\tilde{G} X_s & 0& -\tilde{\Delta}&-\kappa \end{matrix}\right) ,
\end{eqnarray}\normalsize
with  $I\equiv |c_s|^2$, $\tilde{\omega}_m \equiv \omega_m+2g_{_{q}} I$, $\tilde{\Delta}\equiv \Delta+g_{_{l}}x_s +g_{_{q}} x_s^2 $, $X_s=\frac{c_s+c_s^*}{\sqrt{2}}$, $P_s=\frac{c_s-c_s^*}{\sqrt{2}i}$ and
$\tilde{G} \equiv g_{_{l}}  +2 g_{_{q}}x_s$. The solutions are stable only if all the eigenvalues of the matrix $M$ have negative real parts. This can be deduced by applying Routh-Hurwitz criterion \cite{routh} in terms of system parameters \cite{anilkumar}. 

The stability of an OM system can be understood physically as follows. The inclusion of QOC (both magnitude and sign) affect the mechanical spring constant and changes from $K = m\omega_m^2$ to $\tilde{K} =  m\omega_m\tilde{\omega}_m$. While the negative QOC makes the spring softer, the positive QOC makes it stiffer. The modified mechanical spring constant and change in frequency from $\omega_m$ to $\tilde{\omega}_m$\ changes the restoring force, and to balance it, the radiation pressure force given by $F_{rad}=(\frac{\hbar\omega_c}{L})\langle a^\dagger a\rangle$ has to readjust thereafter. In cases where it is not possible to achieve this, the system becomes unstable \cite{anilkumar}. { The present letter examines the limits of achievable force sensitivity using negative QOC. In such cases the system has to satisfy the condition $( \frac{\tilde{\omega}_m}{\omega_m}=1+\frac{2g_qI}{\omega_m})  > 0$ for it to be physical and stable under the Routh-Hurwitz criterion.}

\subsection*{Estimation of force noise spectrum}
\par Let us now consider an external force $\mathcal{F}_{ext}$ acting on the mechanical oscillator as shown in Fig.1. The dimensionless classical force, $F_{ext}=\mathcal{F}_{ext}/\sqrt{\hbar{m}{\omega_m}{\gamma_m}}$, is coupled linearly in position to the mechanical oscillator, and enters the quantum Langevin equations as $\xi'(t)=\xi(t)+F_{ext}(t)$. This shifts the position and changes the effective length of the cavity, causing a variation in the optical phase measured outside the cavity. The phase quadrature can be detected using  homodyne or heterodye techniques. The expression for output phase quadrature can be obtained by expressing the fluctuation equations in frequency domain. Using the definition of Fourier transform, $\mathcal{F}(\omega)=\frac{1}{2\pi}\int_{-\infty}^\infty\mathcal{F}(t)e^{-i\omega t}dt$ and $[\mathcal{F}^\dagger(\omega)]^\dagger=\mathcal{F}(-\omega)$, together with the standard input-output relations \cite{aspelmeyer} { $ \delta P_{out}(\omega)= \sqrt{2\kappa}\delta P (\omega)-\delta  P_{in}(\omega)$}, we get 
\small
\begin{eqnarray}
\delta P_{out}(\omega)&&\hspace{-0.5cm}=\frac{\sqrt{2\kappa}}{D(\omega)}\left[P_x(\omega)\delta x_{in}(\omega)+(P_p(\omega)-D(\omega)/\sqrt{2\kappa})\delta p_{in}(\omega)\right.\nonumber\\&&\left.+P_{\xi}(\omega)\left(\xi(\omega)+F_{ext}\right)\right]
\end{eqnarray}
\normalsize where \small
\begin{subequations}
\begin{eqnarray}
&&P_x(\omega)=-\sqrt{2 \kappa} \left(\omega_m \tilde{G}^2X_s^2 +\tilde{\Delta}(\omega^2+i \gamma_m \omega- \omega_m \tilde{\omega}_m)\right),\\
&&P_{p}(\omega)=-\sqrt{2 \kappa} \left(\omega_m \tilde{G}^2X_sP_s-(\kappa-i\omega)(\omega^2+i \gamma_m \omega- \omega_m \tilde{\omega}_m)\right),\nonumber \\\\
&&P_\xi(\omega)=\tilde{G}\omega_m \sqrt{\gamma_m}\left((\kappa - i \omega)X_s+P_s\tilde{\Delta} \right),\\&& D(\omega)= \left((\kappa -i \omega)^2+\tilde{\Delta}^2\right)\left( \omega^2+i \gamma_m \omega- \omega_m \tilde{\omega}_m)\right)+2\tilde{G}^2I \tilde{\Delta}\omega_m.  \nonumber\\
\end{eqnarray}\label{b2}
\end{subequations}\normalsize 
 To obtain $F_{ext}$, the measured phase quadrature given by Eq(4) can be rewritten and re-scaled as 
\begin{equation}
    \delta P_{out}(\omega)\frac{D(\omega)}{\sqrt{2\kappa}P_{\xi}(\omega)}=F_{N}(\omega)+F_{ext}(\omega),
    \end{equation} where $F_{N}(\omega)$  is the added force noise. The sensitivity of force measurements is estimated by quantifying the spectral density of added noise as:  
\small
\begin{eqnarray}
&&S_{FF}(\omega)=\frac{1}{4\pi}\int e^{-i(\omega+\Omega) t} \langle  F_N(\omega) F_N(\Omega) +F_N(\Omega) F_N(\omega)\rangle d\Omega, \nonumber\\
&&=\frac{2K_BT}{\hbar\omega_m}+\frac{\left(|P_{x}(\omega)|^2+|P_{p}(\omega)-D(\omega)/\sqrt{2\kappa}|^2\right)}{2|P_{\xi}(\omega)|^2}.
\end{eqnarray}\label{a16}
\normalsize

The noise spectrum is a dimensionless quantity and to convert it into force noise spectral density in the units of $N^2/Hz$, we need to multiply by a scalar factor $\hbar m\omega_m\gamma_m$. Therefore the noise spectrum of the OM system is $ S'_{FF}(\omega)= \hbar m \omega_m\gamma_m S_{FF}(\omega)$. 
In order to simplify Eq.(7), we chose optimal case of effective detuning i.e. $\tilde{\Delta}=0$ and by considering measurements confined to $\kappa \gg \omega$ i.e. $(\kappa+ \omega)^2 \sim \kappa^2$ and $I\equiv |c_s|^2=\varepsilon^2/\kappa^2$,  we have
\small
\begin{equation}
    S_{FF}(\omega)_{\tilde{\Delta}=0}=\frac{2K_BT}{\hbar\omega_m}+\frac{\zeta}{2\gamma_m}\left(\frac{\omega_m}{\tilde{\omega}_m}\right)^2+\frac{1}{2\zeta\gamma_m|\tilde{\chi}_m|^2}\left(\frac{\tilde{\omega}_m}{\omega_m}\right)^2
\end{equation}
\normalsize
 where $\tilde{\chi}_m=\omega_m/\left(\omega^2+i\gamma_m\omega-\omega_m\tilde{\omega}_m\right)$  and   $\zeta=4g_{_{l}}^2I/\kappa$. 
 
The first term represents thermal noise which is independent of input power {$(\mathcal{P})$} and depends on temperature (T) adds a flat noise to the force spectrum and there by it reduces the total force sensitivity.  While second term indicates backaction noise  which directly depends on $\zeta$  and inversely proportional to normalized modified spring constant $\left(\tilde{\omega}_m/\omega_m\right)$, the third term represents shot noise that is inversely proportional to backaction noise scaled with the mechanical oscillator susceptibility $|\tilde{\chi}_m(\omega)|$.
\par In order to identify the advantage of our scheme, we choose to perform a numerical analysis in which we compare our scheme to a conventional OM system. The parameters chosen in our calculations are similar to those used in \cite{vitaliprl}. In all our numerical calculations QOC has been scaled with LOC i.e. $\frac{g_{_{q}}}{g_{_{l}}}$. In the present analysis, negative values of the QOC has been considered to exploit soft mode (soft spring effect) of the OM system to explore classical force sensing. 
 
\par \subsection*{Force noise spectrum with no QOC} When Eq(8) is further simplified with the limiting conditions$\frac{g_{_{q}}}{g_{_{l}}} = 0$ and T = 0, it yields    \cite{tsang,meystre,vitali} 
\begin{equation}
    S_{FF}(\omega)\bigg\rvert_{\tilde{\Delta},{\frac{g_{_{q}}}{g_{_{l}}},T=0}} \equiv \tilde{S}_{FF}(\omega)=\frac{\zeta}{2\gamma_m}+\frac{1}{2\zeta\gamma_m|\chi_m|^2}.
\end{equation}
From Eq. (9) one can understand that the competition between back  action (first term)  noise and  shot noise (second term) have  opposite  dependence on the intensity $I\propto \mathcal{P}$ in $\zeta$, which determines SQL. Minimizing Eq.(9) with respect to power ( or equivalently  $\zeta$)  gives lower bound SQL of force noise spectrum for a given detection frequency $\omega $ as $\tilde{S}_{FF}^{SQL}(\omega)=\left(\gamma_m|\chi_m(\omega)|\right)^{-1}$
where $\chi_m=\omega_m/\left(\omega^2+i\gamma_m\omega-\omega^2_m\right)$. For an  on resonance frequency ($\omega = \omega_m$ ), $\tilde{S}_{FF}^{SQL}$ equals 1. We analyze this numerically in Fig.2 which is shown with a  blue dashed curve, mentioned in the legend as $g_{_{q}}/g_{_{l}}=0$ (in log scale). By plotting Eq(8) as a function of $\omega$ for an input power of $\mathcal{P} = 10 \mu W$, we show that the force noise spectrum response is obtained at $\omega=\omega_m$. Also, we have numerically estimated the minimum power required for a standard OM system to reach SQL using Eq.(9) as shown in Fig.3 with black dashed curve. The plot also consists of both backaction noise and shot noise terms as a function of input power ($\mathcal{P}$) and are represented with red and blue color curves respectively. Since the backaction noise and and shot noise  are equal at $\mathcal{P}= 100 \mu W$, the balanced effect of both the noises limit the force spectrum. It is here that the SQL is said to be reached i.e. equal to 1. This achieved value of SQL is also shown in Fig.2 as dashed  black line. The SQL is not a fundamental and unsurpassable limit, but it is an important reference standard in measurement sensitivity where quantum limits starts to prevail over technical limits in weak force sensing. 

\begin{figure}[t]
       \hspace*{-0.3cm} \includegraphics[scale=.425]{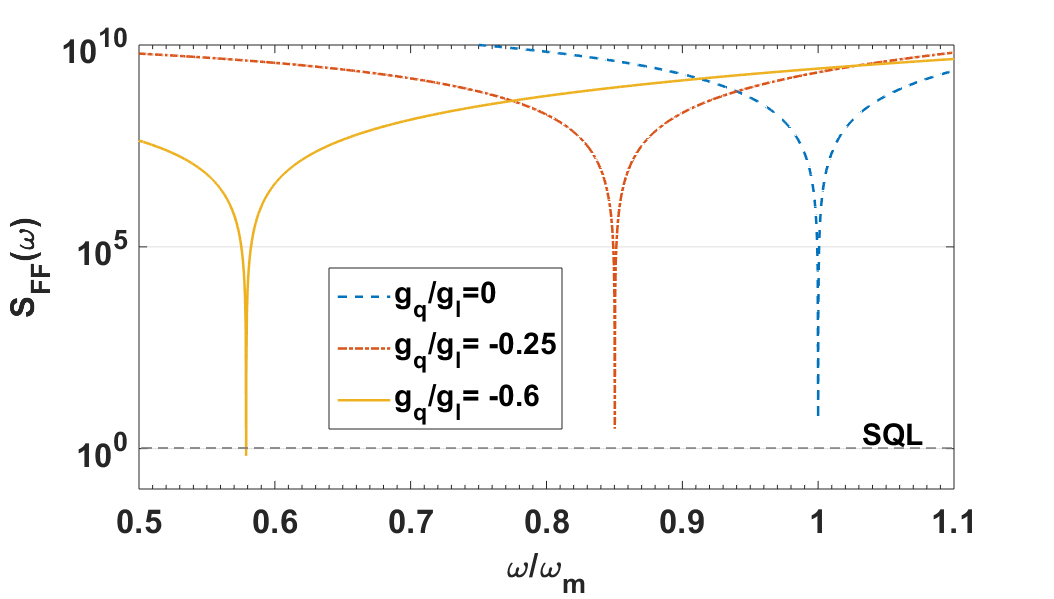}
        \caption{Force noise spectrum $S_{FF}(\omega)$ from Eq.(8) varied as function of $\omega$ and plotted for various values of  negative QOC ($\frac{g_q}{g_l}$) as shown in the legend. The black dashed horizontal line shows the SQL. The OM system is driven by an input laser power $\mathcal{P} =10 \mu W $ of wavelength 810 nm and other system parameters are  $\omega_m/2\pi$ = 10 MHz, $\gamma_m/2\pi$= 100 Hz, $g_{_{l}}/2\pi$ = 215 Hz, $\kappa/2\pi$ = 500 MHz  and  $T=0.$}
    \end{figure}

\begin{figure}[t]
        \centering
        \includegraphics[scale=.40]{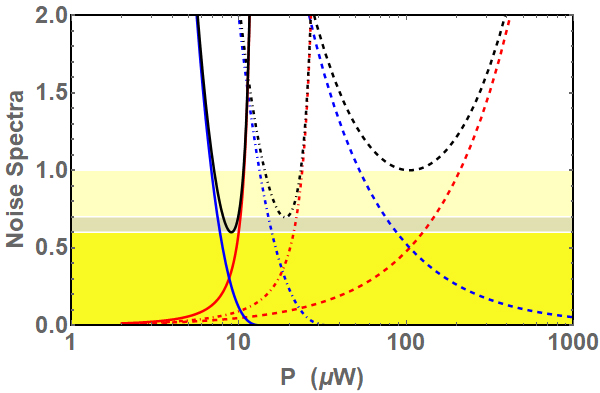}
        \caption{Depicts various noise spectra such as force noise $S_{FF}(\omega)$ from Eq.(8) (Black curves), backaction noise (Red curves) and phase noise (Blue curves) varied as a function of input laser power ($\mathcal{P}$) for various values of QOC ($\frac{g_{_q}}{g_{_l}}$) as 0 (dashed curves), -0.25 (dot dashed curves) and -0.6 (solid curves)  respectively. The regions below SQL are colored for making the data visually appealing. The system parameters as same as mentioned in Fig 2. }
        \label{fig:my_label}
    \end{figure}
    
    \begin{figure}[t]
    \centering
    \includegraphics[scale=0.5]{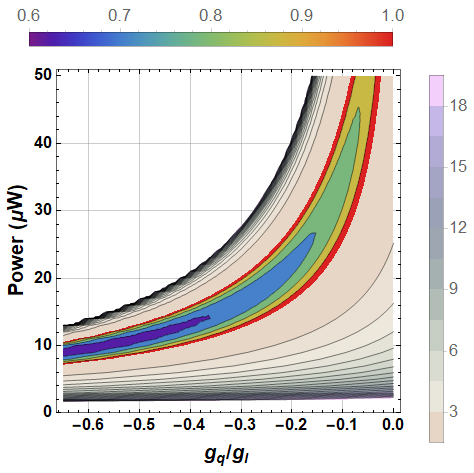}
    \caption{ Force noise spectrum $S_{FF}(\omega)$ (from Eq.(8)) evaluated at various values of soft modes ($\omega = \sqrt{\omega_m\tilde{\omega}_m}$) varied as a function of input power ($\mathcal{P}$) and negative QOC $(g_{_{q}}/g_{_{l}})$. The rainbow colour region shows the enhanced sensitivity beyond SQL (the legend is shown on top of the figure). The gradient purple colour region indicates the $S_{FF}(\omega)> 1$ (the legend is shown on right side of the figure). The white colour region depicts the unstable state of the system.} 
    \end{figure}
 \subsection*{Force noise spectrum with QOC}
\par The inclusion of negative QOC in a regular OM system gives rise to an intensity dependent soft mode and these soft modes changes the mechanical resonance frequency from $\omega_m $ to $\sqrt{\omega_m\tilde{\omega}_m} $. This is also reflected in the mechanical susceptibility i.e. $\tilde{\chi}_m$ appearing in the force noise spectrum of Eq (8). The influence of QOC on force noise spectrum is numerically calculated in Fig. 2 for  QOC values $( g_{_{q}}/g_{_{l}})$ = { -0.25} (red dot-dashed curve),{  -0.6} (orange solid curve) and plotted as a function of $\omega$ for an input power $\mathcal{P} = 10 \mu W$. It is evident that the presence of QOC modifies the  force noise spectrum and for an appropriate value of QOC($(g_{_{q}}/g_{_{l}})={  -0.6}$), it surpasses the SQL. It is also clear that the force noise response is now at the modified mechanical frequency, rather than at $\omega_m$.

The enhancement of force sensitivity can be understood as follows.
  Under the soft mode, the mechanical oscillator motion is nonlinear (from Eq(2)) and thereby more number of photons are accommodated in the cavity. This changes OM interaction from $g_{_{l}}$ to $\tilde{G}$ which is a function of circulating intensity $I\equiv |c_s|^2$ of the optical mode. Therefore $\tilde{G} \equiv g_{_{l}} + 2 g_{_{q}} x_s$ can be written as $ g_{_{l}}\omega_m/\tilde{\omega}_m$ using Eqn.(2).  
 This causes the back-action noise to be a non-linear function of the input power in Eq.(8) as in comparison to Eq.(9). By examining Eq(8) we can see that, in the soft mode regime, redefining $\zeta'=\zeta \left(\frac{\omega_m}{\tilde{\omega}_m}\right)^2$  resembles Eq.(9) at $T=0$, albeit in terms of  $\zeta'$. 

\par We illustrate this by plotting  force noise spectrum and individual added noise terms using Eq.(8) in Fig.3 as a function of  input power ($\mathcal{P}$). The colour code used is same as described above as in the case of no QOC. Fig.3 describe plots for various values of softness ($g_{_{q}}/g_{_{l}}$) with { -0.25} (dot dashed curves) and { -0.6} (solid curves). While the back-action noise turns out to be non-linear with its slope increasing rapidly with respect to the power of the driving field, the shot noise slope decreases rapidly in the soft mode operation of the OM system.  The cumulative change in both backaction noise and shot noise at modified mechanical response$\sqrt{\omega_m\tilde{\omega}_m}$ and $\zeta'$ leads to an overall reduction in the total added noise beyond SQL at lower power, shown in black color. 
This can also be analytically derived by minimizing Eq.(8) with respect to $\zeta'$ that yields:
 \begin{eqnarray}
    &  S^{g_{_{q}}}_{FF}(\omega)=\frac{1}{\gamma_m|\tilde{\chi}_m(\omega)|}, \\
  &   \therefore S_{FF}(\omega)=\left\lbrace \begin{array}{cc}
          \tilde{S}^{SQL}_{FF}(\omega)= 1& \omega=\omega_m,\qquad g_{_{q}}/g_{_{l}}=0\\ S^{g_{_{q}}}_{FF}(\omega)<1& \omega=\sqrt{\omega_m\tilde{\omega}_m}, \quad g_{_{q}}/g_{_{l}}<0
     \end{array} \right..
 \end{eqnarray}
It is evident from Fig. 3 that, as softness increases, the minima of the total added noise moves towards lower power with a value increasingly lower than the SQL. 

\par Since from Eq.(10), we see that, $S_{FF}^{g_q}(\omega)$ is a function of both negative QOC and input power. One has to optimize  both these two parameters to obtain maximum achievable force sensitivity. We therefore plot a contour diagram of $S_{FF}(\omega)$,  as a function of input power and negative QOC evaluated at $\omega=\sqrt{\omega_m \tilde{\omega}_m}$ shown in Fig.4. Here, the values of $S_{FF}(\omega)$> 1  is shown in a gradient purple color, while the rainbow color region depicts the maximal level of force sensitivity that surpasses the SQL.  Also, the figure displays a lower value of total added noise much less than 1 at higher values of negative QOC, which can be attained at lower powers. For example, with an input power (12 $\mu$W) and QOC value (-0.45$g_{_{l}}$), one can easily beat SQL. The total added force noise here is $\sim$0.6 which corresponds to an enhancement in the sensitivity by 40 \% at a power (12 $\mu$W) $\sim$10 times lower than the power required (100 $\mu$W) to achieve SQL in the conventional system (no QOC). The white color region in the Fig.4 shows the unstable region of the system and therefore increasing the softness leads to larger unstable regions in the parameter space of $(\mathcal{P},- g_{_{q}}/g_{_{l}} )$, making it more inaccessible to experimentalists. 

 The QOC values chosen in our numerical calculation are arbitrary and are merely  for a good visibility in showing the effect of QOC on force sensitivity. By choosing much higher values of QOC and very lower power such that $\tilde{\omega}_m>0$ ensures better sensitivities. Also from Eq.(10) we can see that the force sensitivity is dependent purely on the experimental parameters and therefore it is determined by the experimental system. However, in reality i.e. T$\neq$ 0 the achievable force sensitivity is limited by the thermal noise that is proportional to the temperature. For a dilute refrigeration temperature of T= 1 mK, the contribution of thermal noise is 2.068 which ultimately reduces the force sensitivity in a conventional OM system to $\tilde{S}_{FF}(\omega_m) \sim$ 3. However, the presence of negative QOC can lower this further till it is limited by only thermal noise.    

Recent theoretical proposals beating SQL involve CQNC \cite{tsang} demand additional degrees of freedom (DOF) like  OPA \cite{heurs} , ultra-cold atoms \cite{meystre} and hybrid-systems with injecting squeezed vacuum noise \cite{vitali}, complicating the conventional OM architecture.  Our proposal achieves it not by cancelling back-action noise, but by making it a non-linear function of intensity, requiring no additional DOF as in existing experimental platforms like electro-mechanical \cite{schwab,sorensen}, OM  \cite{flowers} and micro disk- cantilevers \cite{jpdavis}. However  progress has to be made to achieve higher ratios of $g_{_{q}}/g_{_{l}}$ to realise our scheme experimentally. As an alternative, hybrid systems like \cite{barker} wherein a nanosphere levitated in a hybrid electro-optical trap could tune $g_{_{q}}$ and $g_{_{l}}$ independently depending on the trapping position. But currently it is a  challenge  to find an optimal temperature such that the $|g_{_{q}}/g_{_{l}}| >0.1$ and the associated thermal noise doesn't degrade the performance. Experimental search for such a system would prove advantageous not only for precision metrology as the current letter suggests but also for realising macroscopic quantum effects.

\medskip
\noindent \textbf{Acknowledgements}  USS was supported in part
at the Technion by a fellowship of the Israel Council for Higher Education and by a Technion fellowship. MAK acknowledges financial support from quantum information wing, AIFI Technologies LLC, UAE.

\medskip

\noindent\textbf{Disclosures.} The authors declare no conflicts of interest.

\bibliography{ref}

\bibliographyfullrefs{ref}

\end{document}